%% file: ms.tex
\documentclass{aa}

\usepackage{comment}
\usepackage{lscape}

\usepackage{natbib}
\bibpunct{(}{)}{;}{a}{}{,} 
\usepackage{amssymb}
\usepackage{graphicx}
\usepackage[usenames,dvipsnames]{color}
\newcommand{\plotone}[1]{\includegraphics[width=0.9\textwidth]{#1}} 
\newcommand{\plothalf}[1]{\includegraphics[width=0.45\textwidth]{#1}} 

\begin{document}

\title{Dust grain growth in the interstellar medium of $5<z<6.5$ quasars}
 
\titlerunning{Dust grain growth in the ISM of $5<z<6.5$ QSOs}
\authorrunning{Micha{\l}owski et al.}

\author{Micha{\l}~J.~Micha{\l}owski\inst{1}
\and
Eric~J.~Murphy\inst{2}
\and
Jens~Hjorth\inst{3}
\and 
Darach~Watson\inst{3}
\and
Christa~Gall\inst{3}
\and
James~S.~Dunlop\inst{1,4}
	}

\institute{
Scottish Universities Physics Alliance, Institute for Astronomy, University of Edinburgh, Royal Observatory, Edinburgh, EH9 3HJ, UK
\and
Spitzer Science Center, MC 314-6, California Institute of Technology, Pasadena, CA 91125, USA
\and
Dark Cosmology Centre, Niels Bohr Institute, University of Copenhagen, Juliane Maries Vej 30, 2100 Copenhagen \O, Denmark
\and
Department of Physics \& Astronomy, University of British Columbia, 6224 Agricultural Road, Vancouver, BC V6T 1Z1, Canada
}

\abstract 
{}
{We investigate whether stellar dust sources i.e.~asymptotic giant branch (AGB) stars and supernovae (SNe) can account for dust detected in $5<z<6.5$ quasars (QSOs).} 
{We calculate the required dust yields per AGB star and per SN using the dust masses of QSOs inferred from their millimeter emission and stellar masses approximated as the difference between the dynamical and the H$_2$ gas masses of these objects.} 
{We find that AGB stars are not efficient enough to form dust in the majority of the $z>5$ QSOs, whereas SNe  may~be~able to account for dust in some QSOs. However, they require very high dust yields even for a top-heavy initial mass function.}
{This suggests additional non-stellar dust formation mechanism e.g.~significant dust grain growth in the interstellar medium of at least three out of nine $z>5$ QSOs. SNe (but not AGB stars) may deliver enough heavy elements to fuel this growth.}

\keywords{dust, extinction - 
galaxies: high-redshift - galaxies: ISM - submillimeter: galaxies - 
quasars: general
}

\maketitle

\section{Introduction}
\label{sec:intro}

\input{masstab}

Studies of the extragalactic background light have revealed that roughly 
 half of the energy emitted in the Universe apart from the CMB is reprocessed by dust  \citep[e.g.][]{hauserdwek01}. 
Thus, understanding the physical processes responsible for the formation of dust throughout cosmic time has important cosmological implications.

Dust can either 
be formed by asymptotic giant branch (AGB) stars  \citep[even at low metallicities;][]{sloan09}, or supernovae (SNe). 
Alternatively, the bulk of the dust mass accumulation may occur in the interstellar medium (ISM) on dust seeds produced by stars. This process 
can successfully explain 
gas depletions in the Milky Way \citep{draine79,dwek80,draine90,draine09}, along with the dust masses of the LMC \citep{matsuura09} and a $z\sim6.42$ quasar \citep[QSO;][]{dwek07}.

Theoretical works have shown that an AGB star and a SN produce up to $\sim4\mbox{}\times10^{-2}\,M_\odot$ \citep{morgan03,ferrarotti06} and $\sim1.32\,M_\odot$ \citep{todini01,nozawa03} of dust, respectively. 
However, for the case of SN dust, only $\lesssim0.1\,M_\odot$ of the dust actually survives in the associated shocks  \citep{bianchi07,cherchneff10}. 

The dust  in the Milky Way was predominantly formed by evolved 
stars with only a minor SN contribution \citep{gehrz89}, but individual SNe may form significant amounts  of dust.
Submillimeter observations of the SN remnants \object{Cassiopeia A} \citep{dunne03,dunne09casA} and Kepler \citep{morgan03b,gomez09} have revealed as much as $\sim1\,M_\odot$ of freshly formed dust, but 
these results are controversial 
\citep{dwek04,krause04,gomez05,wilson05, blair07,sibthorpe09,barlow10}.  Dust yields for other SNe
are typically in the range 
$10^{-3}$--$10^{-2}\,M_\odot$
 \citep{green04,borkowski06,sugerman06,ercolano07,meikle07,rho08,rho09,kotak09,lee09,sakon09,sandstrom09,wesson09}.

The situation is even more complex at high redshifts. 
\citet{dwek07} claimed that only SNe can produce dust on timescales $<1$ Gyr, but 
\citet{valiante09} showed that 
 AGB stars 
 dominate dust production over SNe as early as $150$--$500$ Myr after the onset of star formation. \citet{michalowski10smg4} concluded that in three out of six $4<z<5$ submillimeter galaxies, only SNe are efficient enough to form dust provided that they have high dust yields.
 This would then be
  suggestive of a significant dust growth in the ISM and/or a top-heavy initial mass function 
  (IMF).

Signatures of  SN-origin dust have been claimed 
in the extinction curves of a $z\sim6.2$ QSO (\citeauthor{maiolino04} \citeyear{maiolino04}; see also \citeauthor{gallerani10} \citeyear{gallerani10}) and of two gamma-ray burst host galaxies, one at $z\sim6.3$ (\citeauthor{stratta07b} \citeyear{stratta07b}, but this result was undermined  by \citeauthor{zafar10}  \citeyear{zafar10}) and one at $z\sim5$ \citep{perley09}.

The objective of this paper is to investigate if SNe and AGB stars are efficient enough to form dust at redshifts $5<z<6.5$ ($1.15$--$0.85$ Gyr after the Big Bang), or if  grain growth in the ISM is required.
We use a cosmological model with $H_0=70$ km s$^{-1}$ Mpc$^{-1}$, and ($\Omega_\Lambda, \Omega_m)=(0.7, 0.3)$.

\section{Methodology}
\label{sec:method}

In order to constrain the dust production efficiency in the early Universe, we selected $z>5$ QSOs detected in both the millimeter continuum 
and CO lines 
allowing estimates to be made of
their dust, gas and dynamical masses 
(Tab.~\ref{tab:mass}).

We calculated dust masses ($M_{\rm dust}$) from the  $1200\,\mu$m data 
(rest-frame $160$--$200\,\mu$m)
using Eq.~5 of \citet{michalowski09} assuming $\beta=1.3$. For three QSOs we adopted the derived dust temperatures ($T_{\rm dust}$; Tab.~\ref{tab:mass}). 
For the rest we assumed the average of these estimates $T_{\rm dust}=50$~K.
We assumed the mass absorption coefficient $\kappa_{1200\,\mu{\rm m}}=0.67\,{\rm cm}^2 {\rm g}^{-1}$, a conservatively high value \citep[cf.][]{alton04} resulting in systematically low $M_{\rm dust}$.

In order to explore the impact of systematic uncertainties on $M_{\rm dust}$, we also assumed $\beta=2.0$ \citep[see][]{dunne00,dunne01,vlahakis05}. This gives $M_{\rm dust}$ smaller by a factor of $\sim3.75$  \citep[see Fig.~3 of][]{michalowski10smg}. Changing $T_{\rm dust}$ to a very high value of $80$ K 
\citep[compare with Fig.~2 of][]{michalowski08}, i.e., an upper bound for other QSOs \citep{haas98,benford99,leech01,priddey01,knudsen03,beelen06,aravena08,leipski10},
decreases the $M_{\rm dust}$ by a factor of $\sim2.3$. Hence, we also assumed $(T_{\rm dust},\beta)=(80,2.0)$. This results in strict lower limits on $M_{\rm dust}$ smaller by a factor of $3.75\times2.3=8.6$. However, this very conservative assumption is only chosen to illustrate an extreme limit on $M_{\rm dust}$.  
It is not likely that the real values are close to this limit as $T_{\rm dust}$ has been constrained to be below $60$~K for four out of nine QSOs in our sample with good wavelength coverage in the infrared (Tab.~\ref{tab:mass}).

{Similar to \cite{wang10}, we assume that the stellar masses ($M_*$) of the QSO host galaxies can be approximated as the difference between the dynamical ($M_{\rm dyn}$; i.e.~total) and the H$_2$ gas masses ($M_{\rm gas}$).}
 The true values of $M_*$ are lower, unless QSOs harbour very little atomic  gas (\ion{H}{I}).  Given the significant uncertainties in the conversion from CO line strength to $M_{\rm gas}$, we also performed the calculations with the upper limit setting $M_*$ equal to $M_{\rm dyn}$.

The inclination angle of the gas disk, $i$, was adopted to be $65^\circ$ for QSO 6 \citep{walter04} and $40^\circ$ for the others \citep{wang10}. The latter assumption is a major source of uncertainty in our analysis and is discussed below.

We calculated the dust yields per  AGB star and per SN (amount of dust formed in ejecta of one star) required to explain the inferred dust masses in the $z>5$ QSOs as described in \citet{michalowski10smg4}. The number of stars with masses between $M_0$ and $M_1$ in the stellar population with a total mass of $M_*$ was calculated as $N(M_0$--$M_1)=M_* \int_{M_0}^{M_1} M^{-\alpha} dM / \int_{M_{\rm min}}^{M_{\rm max}} M^{-\alpha}  M dM$. We adopted an IMF with $M_{\rm min}=0.15$, $M_{\rm max}=120\,M_\odot$, and a slope $\alpha=2.35$ \citep[][or $\alpha=1.5$ for a top-heavy IMF]{salpeter}. 
The average dust yield per  star is $M_{\rm dust} /  N(M_0$--$M_1)$.

\section{Results and Discussion}
\label{sec:results}

\begin{figure*}
\begin{center}
\plotone{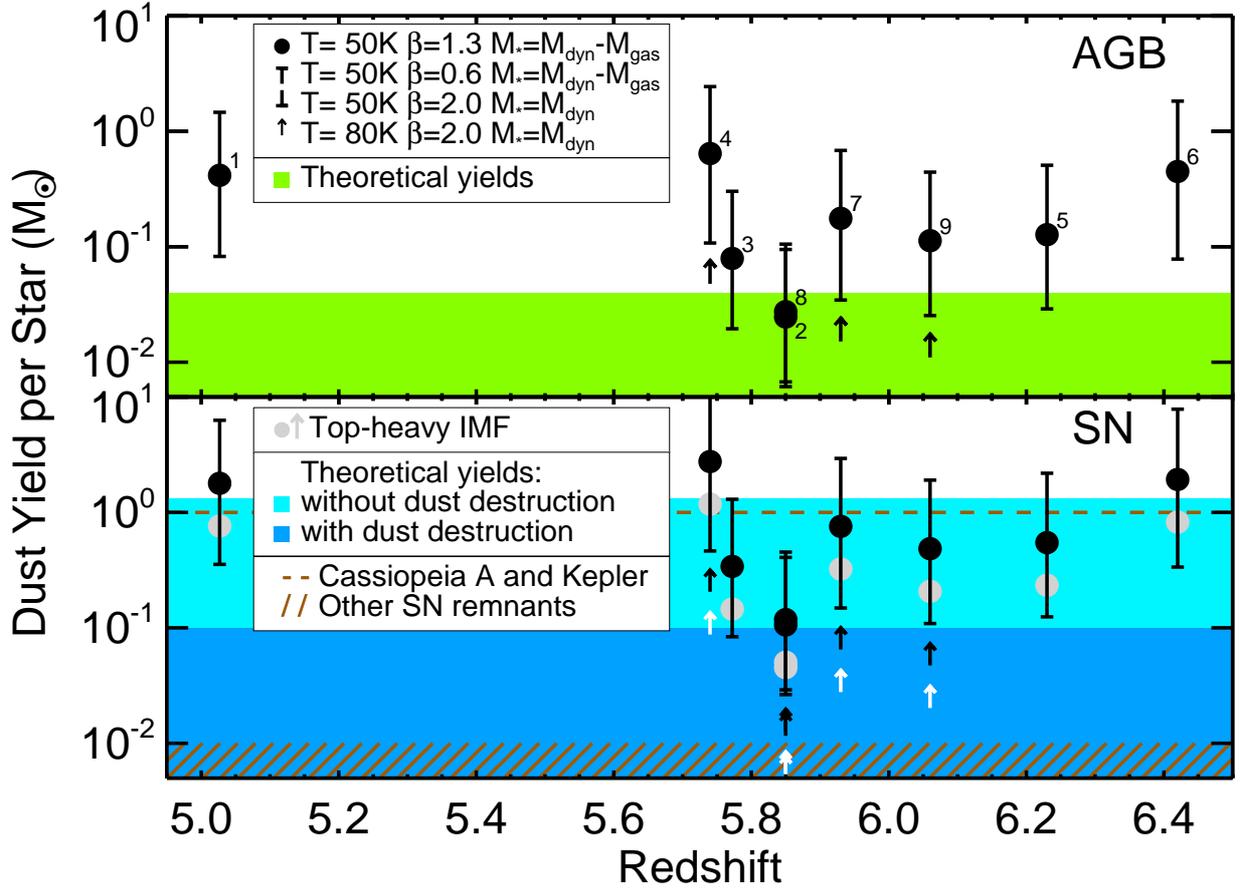}
\end{center}
\caption{Dust yields per AGB star ({\it top}) or per SN ({\it bottom}) required to explain dust in the $z>5$ QSOs. 
For reasonable assumptions on the dust properties, AGB stars are not efficient enough  and SNe would need to be unfeasibly efficient to form dust in these sources suggesting rapid grain grown in the ISM is likely to be responsible for the large dust masses.
{\it Circles}: the best estimates of the required dust yields with error bars reflecting the uncertainty of $\beta$ and $M_*$.
Numbers indicate the QSOs as in Tab.~\ref{tab:mass}.
{\it Arrows}: strict and unlikely lower limits with very high $T_{\rm dust}$ and $\beta$ shown where data allow it (Tab.~\ref{tab:mass}).
{\it Gray symbols} indicate that a top-heavy IMF was adopted.
{\it Dashed line} and {\it diagonal lines}: the dust yields derived for  \object{Cassiopeia A}, 
Kepler
($\sim1\,M_\odot$)
and other SN remnants 
($\sim10^{-3}$--$10^{-2}\,M_\odot$),
respectively. 
{\it Green area}: theoretical dust yields for AGB stars 
($\lesssim4\cdot10^{-2}\,M_\odot$).
{\it Light blue} and {\it blue areas}: theoretical SN dust yields without 
($\lesssim1.32\,M_\odot$)
and with dust destruction implemented 
($\lesssim0.1\,M_\odot$), respectively.
}
\label{fig:dust}
\end{figure*}

\input{dustyieldstab}

First, we consider a single dust producer i.e.~assume that  dust in the $z>5$ QSOs was produced by either AGB stars or SNe.
The required  dust yields per  AGB star and per SN 
are listed in Tab.~\ref{tab:dustyields} and shown in Fig.~\ref{fig:dust} as a function of redshift. 
Circles correspond to reasonable estimates of $T_{\rm dust}$, $\beta$ and $M_*$, whereas other values are shown to quantify the impact of the systematic uncertainties 
(error bars extend down to the reasonable lower limits, whereas arrows represent strict and unlikely lower limits).

Except for 
QSOs 2 and 8 the required yields for AGB stars exceed the theoretically allowed maximum values (green area) by a factor of $2$--$15$. The yields remain too high even for $M_*=M_{\rm dyn}$ and $\beta=2$. They are consistent (though at the high end) with the theoretical predictions only under the unrealistic assumption of $(T_{\rm dust},\beta)=(80,2.0)$.  
Using the $T_{\rm dust}$ limits (Tab.~\ref{tab:mass}), we can robustly rule out a significant contribution of AGB stars to the dust formation in five out of nine QSOs  (1, 3,  4, 5 and 6) and  rule out their contribution in  QSOs 7 and 9, unless their emission is dominated by hot ($\sim80$ K) dust.

Therefore AGB stars are not efficient enough to form dust in the majority of the $z>5$ QSOs.
This contradicts the claim of \citet{valiante09} that $\sim80$\% of dust in 
QSO 6
was created by AGB stars. 
The disagreement can be traced to the fact that
they assumed $M_*\sim10^{12}\,M_\odot$, exceeding the $M_{\rm dyn}$ by a factor of $\sim15$.

For only two
QSOs (2 and 8) are
the required SN dust yields marginally within the theoretically predicted limits with dust destruction implemented (dark blue area on Fig.~\ref{fig:dust}). For the remaining seven QSOs, one would need to assume unrealistically high $T_{\rm dust}$ {\it and} steep spectral slopes {\it and} in some cases an IMF   more top-heavy than the Salpeter IMF.

For these seven QSOs
(including QSO 5 for which \citeauthor{maiolino04} \citeyear{maiolino04} claimed SN-origin dust)
the required SN dust yields are within the theoretical limits without dust destruction (light blue area) and the values observed for SN remnants \object{Cassiopeia A}  and Kepler (dashed line).

\begin{figure}
\begin{center}
\plothalf{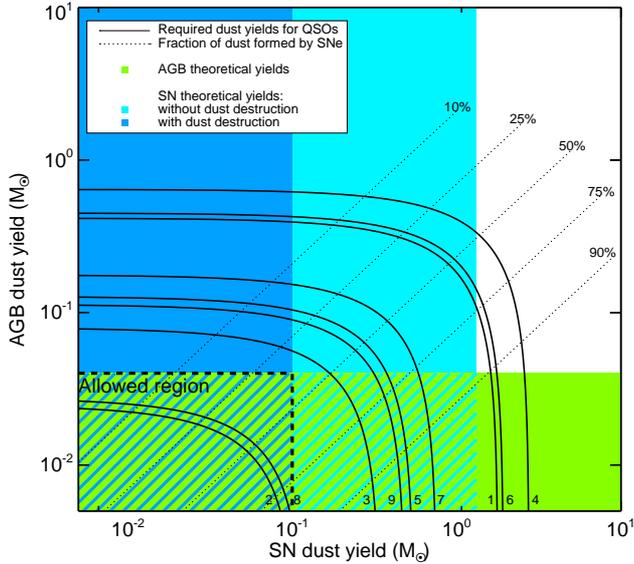}
\end{center}
\caption{The relation of the required dust yields per AGB star and per SN for different fractions of dust formed by SNe (shown as {\it dotted lines}). This is a combination of panels in Fig.~\ref{fig:dust} relaxing the assumption that only AGB stars or only SNe produced dust in the $z>5$ QSOs. The theoretically allowed regions of dust yields are shown as in Fig~\ref{fig:dust}. {\it Hashed region outlined by the dashed line} corresponds to the the allowed region, where the dust yields for both AGB stars and SNe  are within theoretical limits (with the dust destruction implemented).   The {\it solid lines} correspond to the $z>5$ QSOs numbered as in Tab.~\ref{tab:mass}. If higher fraction of dust is attributed to SNe then the QSOs move towards bottom-right corner.   The combined effort of AGB stars and SNe can explain dust in QSO 2 and 8, but not in QSO 1, 4 and 6. Dust in QSOs 3, 5, 7 and 9 may have been formed by these stellar sources, but only if little dust is destroyed in SN shocks and that SN account for more than $50$--$75$\% of dust in these QSOs. 
}
\label{fig:2}
\end{figure}

We checked that allowing AGB stars to form only a fraction of dust in the $z>5$ QSOs and assigning the rest to SNe 
may have an impact on our conclusions for only four out of nine QSOs (3, 5, 7 and 9).
This is illustrated in Fig.~\ref{fig:2}, where we show the required dust yields assuming different fractions of dust attributed to SNe. Solid lines represents the required yields for the QSOs. An increase in the fraction of SN dust corresponds to moving towards bottom-right corner (i.e.~higher SN yields and lower AGB yields are required). If a curve corresponding to a QSO crosses the hatched region, corresponding to the allowed yields for both AGB stars and SNe, then these stellar objects can account for dust in this QSO. Hence we conclude similarly as before, that combined AGB stars and SNe   are efficient enough to form dust in QSOs 2 and 8 and are not efficient enough for QSO 1, 4, and 6. The stellar dust producers may account for dust in QSOs 3, 5, 7  and 9, but only if very little dust is destroyed in SN shocks (light blue area on Fig.~\ref{fig:2}). For these cases, SNe should be responsible for more than $50$--$75$\% of dust in these QSOs.  Alternatively, the required dust yields for these four QSOs can be reconciled with theoretical expectations with dust destruction implemented (dark blue area on Fig.~\ref{fig:2}) if we assume a high value of $\beta=2$.

We stress that our results are sensitive to the assumed gas disk inclinations. The required AGB and SN dust yields for individual QSOs decrease to theoretically allowed values (with dust destruction) for inclinations lower than $5$--$20^\circ$. It is however unlikely that all our QSOs exhibit such low inclination (e.g.~\citeauthor{polletta08} \citeyear{polletta08} did not find any preferred inclination for luminous QSOs). At least this is not the case for QSO 6 with a measured inclination of $\sim65^\circ$.

Moreover, our derived required dust yields should be corrected towards lower values if {\it i}) the gas disk radius of QSOs is larger than $2.5$ kpc assumed by \citet[][then the dynamical mass would be larger]{wang10}, or {\it ii}) the stellar component is more extended than the gas disk (then our upper limit on $M_*$ equal to $M_{\rm dyn}$ would apply only to the stellar component distributed within the extent of the gas disk). 

It is however unlikely that these conditions are fulfilled in our sample. Using the high-resolution CO line observations, the sizes of the gas disks have been constrained for QSO 1 \citep[$<3$ kpc;][]{maiolino07}, QSO 5 \citep[$2.2\times5.0$ kpc;][]{wang10} and QSO 6 \citep[$2.5$ kpc;][]{walter04}. Moreover, the star-forming gas of QSO 6 has been found to be distributed within a radius of  0.75 kpc   \citep{walter09}.

There is no estimate of the extent of the stellar component of the $z>5$ QSOs, but at redshifts $\sim0-3$ QSOs are typically hosted in $\lesssim3$ kpc galaxies \citep{ridgway01,veilleux09},  consistent with a value of $2.5$ kpc assumed by \citet{wang10}.

Hence, we conclude that, unless the inclinations are biased low  or the extent of stellar component are significantly larger than $2.5$ kpc, both AGB stars and SNe would have to form unfeasibly large amounts of dust to account for dust present in the $z>5$ QSOs. This may be taken as an indication of another (non-stellar) dust source in these objects, 
e.g., significant grain growth in the  ISM \citep[e.g.,][]{draine79,dwek80,draine09} on the dust seeds produced by SNe or possibly AGB stars. 
Assuming that star formation in these QSOs began at $z\sim10$, a timescale for {\it in situ} grain growth of a few$\mbox{}\times10$ Myr \citep{draine90,draine09,hirashita00,zhukovska08} is $\la$10\% of the available time, suggesting ample time for grain growth in the ISM to explain the observed dust masses.    

Do stellar sources deliver enough additional heavy elements (not incorporated in dust) necessary for grain growth? The majority of heavy elements produced by AGB stars are already bound in dust grains \citep[yields of carbon and other heavy elements are $\lesssim3.5\times10^{-2}\,M_\odot$;][]{morgan03}. On the other hand, a SN produces as much as $\lesssim1\,M_\odot$ of heavy elements \citep{todini01,nozawa03,bianchi07,cherchneff09}, close to the required yields for the $z>5$ QSOs (lower panel of Fig.~\ref{fig:dust}). Hence, even though SNe themselves do not produce enough dust, they may deliver enough  heavy elements to fuel the dust grain growth in the ISM.

\section{Conclusions}
\label{sec:conclusion}

We have derived the  dust yields per AGB star and per SN required to explain observationally determined dust masses in $5<z<6.5$ QSOs. We  find that the yields for AGB stars typically exceed the theoretically allowed values making these objects inefficient to produce dust at high redshifts. SNe could in principle be responsible for dust in some of the QSOs, but with a requirement of high dust yields. 
This advocates for non-stellar dust source e.g.~significant dust grain growth in the ISM of  at least three out of nine QSOs. We argue that SNe deliver enough heavy elements to fuel the dust growth.

\begin{acknowledgements}

We  thank Joanna Baradziej, Karina Caputi, Eli Dwek and our anonymous referee
 for help with improving this paper.
The Dark Cosmology Centre is funded by the Danish National Research Foundation.

\end{acknowledgements}


\input{ms.bbl}

\end{document}

%% file: masstab.tex
\begin{table*}
\caption{Dust, gas and dynamical masses of $z>5$ QSOs \label{tab:mass}}
\centering
\begin{tabular}{clcccccc}
\hline\hline
&    &     & $M_{\rm dust}$ & $M_{\rm gas}$ & $M_{\rm dyn}\sin^2i$ & &  $T_{\rm dust}$\\
No. & QSO & $z$ & $(10^8M_\odot)$  & $(10^{10}M_\odot)$ & $(10^{10}M_\odot)$ & $\frac{M_{\rm gas}}{M_{\rm dust}}$ & (K)\\
\hline
       1&\object{J0338+0021}&5.03&7.1$\pm$0.6&2.2 \tablefootmark{a}&3.0
\tablefootmark{a}&          31 & 45.6 \tablefootmark{d}\\
       2&\object{J0840+5624}&5.85&4.7$\pm$0.9&2.5 \tablefootmark{b}&24.2
\tablefootmark{b}&          53 & \dots\\
       3&\object{J0927+2001}&5.77&7.2$\pm$1.1&1.8 \tablefootmark{b}&11.8
\tablefootmark{b}&          25 & 51.1 \tablefootmark{d}\\
       4&\object{J1044$-$0125}&5.74&2.7$\pm$0.6&0.7 \tablefootmark{b}&0.8
\tablefootmark{b}&          26 & \dots\\
       5&\object{J1048+4637}&6.23&4.3$\pm$0.6&1.0 \tablefootmark{b}&4.5
\tablefootmark{b}&          23 & $<$40 \tablefootmark{e}\\
       6&\object{J1148+5251}&6.42&5.9$\pm$0.7&1.6 \tablefootmark{c}&4.5
\tablefootmark{c}&          27 & 55.0 \tablefootmark{f}\\
       7&\object{J1335+3533}&5.93&3.4$\pm$0.7&1.8 \tablefootmark{b}&3.1
\tablefootmark{b}&          53 & \dots\\
       8&\object{J1425+3254}&5.85&3.3$\pm$0.7&2.0 \tablefootmark{b}&15.6
\tablefootmark{b}&          60 & \dots\\
       9&\object{J2054$-$0005}&6.06&3.4$\pm$0.8&1.2 \tablefootmark{b}&4.2
\tablefootmark{b}&          35 & \dots\\
\hline
\end{tabular}
\tablefoot{The sample includes QSOs detected in their dust continuum and CO line emission \citep{carilli00,carilli01,carilli07,bertoldi03, bertoldi03b,petric03,priddey03,priddey08,walter03,walter04,robson04,beelen06, maiolino07, riechers07,wang07b,wang08,wang08b,wang10,wu09}. We calculated dust masses from the detection of $1200\,\mu$m emission  using Eq.~5 of \citet{michalowski09} assuming $\beta=1.3$. The errors reflect the statistical uncertainties only. For three QSOs we assumed observationally inferred dust temperatures (the last column). For the rest we adopted the average of these estimates $T_{\rm dust}=50$~K. 
\tablefoottext{a}{\citet{maiolino07}.}
\tablefoottext{b}{\citet{wang10}.}
\tablefoottext{c}{\citet{walter04}.}
\tablefoottext{d}{\citet{wang08b}.}
\tablefoottext{e}{\citet{robson04}.}
\tablefoottext{f}{\citet{beelen06}.}
}
\end{table*}

%% file: dustyieldstab.tex
\begin{table*}
\caption{Dust yields per star required to explain dust in $z>5$ QSOs \label{tab:dustyields}}
\centering
\begin{tabular}{lccll cccccccccc}
\hline\hline
& $T_d$ & & & \multicolumn{           9}{c}{Dust Yields ($M_\odot$ Per Star)}\\
\cline{6-          14}
Dust Producer & (K) & $\beta$ & IMF & $M_*$ &        1&       2&       3&       4&       5&       6&       7&       8&       9&Sym\\
\hline
AGB ($2.5$-$8M_\odot$) & 50 & 1.3 & Sal. & $M_{\rm dyn}$-$M_{\rm gas}$
 & 0.42
 & 0.02
 & 0.08
 & 0.64
 & 0.13
 & 0.45
 & 0.18
 & 0.03
 & 0.11
&$\bullet$\\
AGB ($2.5$-$8M_\odot$) & 50 & 2.0 & Sal. & $M_{\rm dyn}$
 & 0.08
 & 0.01
 & 0.02
 & 0.11
 & 0.03
 & 0.08
 & 0.03
 & 0.01
 & 0.03
&$\perp$\\
AGB ($2.5$-$8M_\odot$) & 80 & 2.0 & Sal. & $M_{\rm dyn}$
 & 0.031
 & 0.003
 & 0.009
 & 0.048
 & 0.012
 & 0.039
 & 0.015
 & 0.003
 & 0.011
&$\uparrow$\\
\hline
SN ($8$-$40M_\odot$) & 50 & 1.3 & Sal. & $M_{\rm dyn}$-$M_{\rm gas}$
 & 1.79
 & 0.11
 & 0.34
 & 2.76
 & 0.55
 & 1.93
 & 0.76
 & 0.12
 & 0.48
&$\bullet$\\
SN ($8$-$40M_\odot$) & 50 & 1.3 & Top & $M_{\rm dyn}$-$M_{\rm gas}$
 & 0.76
 & 0.05
 & 0.15
 & 1.18
 & 0.23
 & 0.82
 & 0.32
 & 0.05
 & 0.21
&{\color{Gray} $\bullet$}\\
SN ($8$-$40M_\odot$) & 50 & 2.0 & Sal. & $M_{\rm dyn}$
 & 0.35
 & 0.03
 & 0.08
 & 0.46
 & 0.12
 & 0.34
 & 0.15
 & 0.03
 & 0.11
&$\perp$\\
SN ($8$-$40M_\odot$) & 80 & 2.0 & Sal. & $M_{\rm dyn}$
 & 0.13
 & 0.01
 & 0.04
 & 0.21
 & 0.05
 & 0.17
 & 0.06
 & 0.01
 & 0.05
&$\uparrow$\\
SN ($8$-$40M_\odot$) & 80 & 2.0 & Top & $M_{\rm dyn}$
 & 0.057
 & 0.005
 & 0.016
 & 0.088
 & 0.023
 & 0.072
 & 0.028
 & 0.005
 & 0.020
&{\color{Gray} $\uparrow$}\\
\hline
\end{tabular}
\tablefoot{The IMF is either \citet{salpeter} with $\alpha=2.35$ or top-heavy with $\alpha=1.5$. The $M_*$ column indicates either that stellar mass was assumed to be the difference between the dynamical and gas masses ($M_{\rm dyn}-M_{\rm gas}$)  or that the strict upper limit to the stellar mass equal to the dynamical mass was adopted (see Sec.~\ref{sec:method}). The numbered columns contain the required dust yields for all QSOs in the order given in Table~\ref{tab:mass}. Only their numbers are given for brevity. The last column gives the symbol used on Fig.~\ref{fig:dust}.}
\end{table*}

%% file: ms.bbl
\begin{thebibliography}{86}
\expandafter\ifx\csname natexlab\endcsname\relax\def\natexlab#1{#1}\fi
\expandafter\ifx\csname url\endcsname\relax
  \def\url#1{{\tt #1}}\fi
\expandafter\ifx\csname urlprefix\endcsname\relax\def\urlprefix{URL }\fi

\bibitem[{{Alton} et~al.(2004){Alton}, {Xilouris}, {Misiriotis}, {Dasyra}, \&
  {Dumke}}]{alton04}
{Alton} P.B., {Xilouris} E.M., {Misiriotis} A., {Dasyra} K.M., {Dumke} M.,
  2004, \aap, 425, 109

\bibitem[{{Aravena} et~al.(2008){Aravena}, {Bertoldi}, {Schinnerer}
  et~al.}]{aravena08}
{Aravena} M., {Bertoldi} F., {Schinnerer} E., et~al., 2008, \aap, 491, 173

\bibitem[{{Barlow} et~al.(2010){Barlow}, {Krause}, {Swinyard}
  et~al.}]{barlow10}
{Barlow} M.J., {Krause} O., {Swinyard} B.M., et~al., 2010, \aap, accepted, {\tt
  arXiv:1005.2688}

\bibitem[{{Beelen} et~al.(2006){Beelen}, {Cox}, {Benford} et~al.}]{beelen06}
{Beelen} A., {Cox} P., {Benford} D.J., et~al., 2006, \apj, 642, 694

\bibitem[{{Benford} et~al.(1999){Benford}, {Cox}, {Omont}, {Phillips}, \&
  {McMahon}}]{benford99}
{Benford} D.J., {Cox} P., {Omont} A., {Phillips} T.G., {McMahon} R.G., 1999,
  \apjl, 518, L65

\bibitem[{{Bertoldi} et~al.(2003{\natexlab{a}}){Bertoldi}, {Carilli}, {Cox}
  et~al.}]{bertoldi03}
{Bertoldi} F., {Carilli} C.L., {Cox} P., et~al., 2003{\natexlab{a}}, \aap, 406,
  L55

\bibitem[{{Bertoldi} et~al.(2003{\natexlab{b}}){Bertoldi}, {Cox}, {Neri}
  et~al.}]{bertoldi03b}
{Bertoldi} F., {Cox} P., {Neri} R., et~al., 2003{\natexlab{b}}, \aap, 409, L47

\bibitem[{{Bianchi} \& {Schneider}(2007)}]{bianchi07}
{Bianchi} S., {Schneider} R., 2007, \mnras, 378, 973

\bibitem[{{Blair} et~al.(2007){Blair}, {Ghavamian}, {Long} et~al.}]{blair07}
{Blair} W.P., {Ghavamian} P., {Long} K.S., et~al., 2007, \apj, 662, 998

\bibitem[{{Borkowski} et~al.(2006){Borkowski}, {Williams}, {Reynolds}
  et~al.}]{borkowski06}
{Borkowski} K., {Williams} B., {Reynolds} S., et~al., 2006, \apjl, 642, L141

\bibitem[{{Carilli} et~al.(2000){Carilli}, {Bertoldi}, {Menten}
  et~al.}]{carilli00}
{Carilli} C.L., {Bertoldi} F., {Menten} K.M., et~al., 2000, \apjl, 533, L13

\bibitem[{{Carilli} et~al.(2001){Carilli}, {Bertoldi}, {Rupen}
  et~al.}]{carilli01}
{Carilli} C.L., {Bertoldi} F., {Rupen} M.P., et~al., 2001, \apj, 555, 625

\bibitem[{{Carilli} et~al.(2007){Carilli}, {Neri}, {Wang} et~al.}]{carilli07}
{Carilli} C.L., {Neri} R., {Wang} R., et~al., 2007, \apjl, 666, L9

\bibitem[{{Cherchneff} \& {Dwek}(2009)}]{cherchneff09}
{Cherchneff} I., {Dwek} E., 2009, \apj, 703, 642

\bibitem[{{Cherchneff} \& {Dwek}(2010)}]{cherchneff10}
{Cherchneff} I., {Dwek} E., 2010, \apj, 713, 1

\bibitem[{{Draine}(1990)}]{draine90}
{Draine} B.T., 1990, In: {L.~Blitz} (ed.) The Evolution of the Interstellar
  Medium, vol.~12 of ASP Conf. Series, 193--205

\bibitem[{{Draine}(2009)}]{draine09}
{Draine} B.T., 2009, In: Th.~Henning J.S. E.~Grun (ed.) Cosmic Dust -- Near and
  Far, ASP Conf. Series, {\tt arXiv:0903.1658}

\bibitem[{{Draine} \& {Salpeter}(1979)}]{draine79}
{Draine} B.T., {Salpeter} E.E., 1979, \apj, 231, 438

\bibitem[{{Dunne} \& {Eales}(2001)}]{dunne01}
{Dunne} L., {Eales} S.A., 2001, \mnras, 327, 697

\bibitem[{{Dunne} et~al.(2000){Dunne}, {Eales}, {Edmunds} et~al.}]{dunne00}
{Dunne} L., {Eales} S., {Edmunds} M., et~al., 2000, \mnras, 315, 115

\bibitem[{{Dunne} et~al.(2003){Dunne}, {Eales}, {Ivison}, {Morgan}, \&
  {Edmunds}}]{dunne03}
{Dunne} L., {Eales} S., {Ivison} R., {Morgan} H., {Edmunds} M., 2003, \nat,
  424, 285

\bibitem[{{Dunne} et~al.(2009){Dunne}, {Maddox}, {Ivison} et~al.}]{dunne09casA}
{Dunne} L., {Maddox} S.J., {Ivison} R.J., et~al., 2009, \mnras, 394, 1307

\bibitem[{{Dwek}(2004)}]{dwek04}
{Dwek} E., 2004, \apj, 607, 848

\bibitem[{{Dwek} \& {Scalo}(1980)}]{dwek80}
{Dwek} E., {Scalo} J.M., 1980, \apj, 239, 193

\bibitem[{{Dwek} et~al.(2007){Dwek}, {Galliano}, \& {Jones}}]{dwek07}
{Dwek} E., {Galliano} F., {Jones} A.P., 2007, \apj, 662, 927

\bibitem[{{Ercolano} et~al.(2007){Ercolano}, {Barlow}, \&
  {Sugerman}}]{ercolano07}
{Ercolano} B., {Barlow} M.J., {Sugerman} B.E.K., 2007, \mnras, 375, 753

\bibitem[{{Ferrarotti} \& {Gail}(2006)}]{ferrarotti06}
{Ferrarotti} A.S., {Gail} H.P., 2006, \aap, 447, 553

\bibitem[{{Gallerani} et~al.(2010){Gallerani}, {Maiolino}, {Juarez}
  et~al.}]{gallerani10}
{Gallerani} S., {Maiolino} R., {Juarez} Y., et~al., 2010, \aap, accepted, {\tt
  arXiv:1006.4463}

\bibitem[{{Gehrz}(1989)}]{gehrz89}
{Gehrz} R., 1989, In: {L.~J.~Allamandola \& A.~G.~G.~M.~Tielens} (ed.)
  Interstellar Dust, vol. 135 of IAU Symposium, 445

\bibitem[{{Gomez} et~al.(2005){Gomez}, {Dunne}, {Eales}, {Gomez}, \&
  {Edmunds}}]{gomez05}
{Gomez} H.L., {Dunne} L., {Eales} S.A., {Gomez} E.L., {Edmunds} M.G., 2005,
  \mnras, 361, 1012

\bibitem[{{Gomez} et~al.(2009){Gomez}, {Dunne}, {Ivison} et~al.}]{gomez09}
{Gomez} H.L., {Dunne} L., {Ivison} R.J., et~al., 2009, \mnras, 397, 1621

\bibitem[{{Green} et~al.(2004){Green}, {Tuffs}, \& {Popescu}}]{green04}
{Green} D.A., {Tuffs} R.J., {Popescu} C.C., 2004, \mnras, 355, 1315

\bibitem[{{Haas} et~al.(1998){Haas}, {Chini}, {Meisenheimer} et~al.}]{haas98}
{Haas} M., {Chini} R., {Meisenheimer} K., et~al., 1998, \apjl, 503, L109

\bibitem[{{Hauser} \& {Dwek}(2001)}]{hauserdwek01}
{Hauser} M.G., {Dwek} E., 2001, \araa, 39, 249

\bibitem[{{Hirashita}(2000)}]{hirashita00}
{Hirashita} H., 2000, \pasj, 52, 585

\bibitem[{{Knudsen} et~al.(2003){Knudsen}, {van der Werf}, \&
  {Jaffe}}]{knudsen03}
{Knudsen} K.K., {van der Werf} P.P., {Jaffe} W., 2003, \aap, 411, 343

\bibitem[{{Kotak} et~al.(2009){Kotak}, {Meikle}, {Farrah} et~al.}]{kotak09}
{Kotak} R., {Meikle} W.P.S., {Farrah} D., et~al., 2009, \apj, 704, 306

\bibitem[{{Krause} et~al.(2004){Krause}, {Birkmann}, {Rieke} et~al.}]{krause04}
{Krause} O., {Birkmann} S.M., {Rieke} G.H., et~al., 2004, \nat, 432, 596

\bibitem[{{Lee} et~al.(2009){Lee}, {Koo}, {Moon} et~al.}]{lee09}
{Lee} H., {Koo} B., {Moon} D., et~al., 2009, \apj, 706, 441

\bibitem[{{Leech} et~al.(2001){Leech}, {Metcalfe}, \& {Altieri}}]{leech01}
{Leech} K.J., {Metcalfe} L., {Altieri} B., 2001, \mnras, 328, 1125

\bibitem[{{Leipski} et~al.(2010){Leipski}, {Meisenheimer}, {Klaas}
  et~al.}]{leipski10}
{Leipski} C., {Meisenheimer} K., {Klaas} U., et~al., 2010, \aap, accepted, {\tt
  arXiv:1005.5016}

\bibitem[{{Maiolino} et~al.(2004){Maiolino}, {Schneider}, {Oliva}
  et~al.}]{maiolino04}
{Maiolino} R., {Schneider} R., {Oliva} E., et~al., 2004, \nat, 431, 533

\bibitem[{{Maiolino} et~al.(2007){Maiolino}, {Neri}, {Beelen}
  et~al.}]{maiolino07}
{Maiolino} R., {Neri} R., {Beelen} A., et~al., 2007, \aap, 472, L33

\bibitem[{{Matsuura} et~al.(2009){Matsuura}, {Barlow}, {Zijlstra}
  et~al.}]{matsuura09}
{Matsuura} M., {Barlow} M., {Zijlstra} A., et~al., 2009, \mnras, 396, 918

\bibitem[{{Meikle} et~al.(2007){Meikle}, {Mattila}, {Pastorello}
  et~al.}]{meikle07}
{Meikle} W.P.S., {Mattila} S., {Pastorello} A., et~al., 2007, \apj, 665, 608

\bibitem[{{Micha{\l}owski} et~al.(2010{\natexlab{a}}){Micha{\l}owski},
  {Hjorth}, \& {Watson}}]{michalowski10smg}
{Micha{\l}owski} M., {Hjorth} J., {Watson} D., 2010{\natexlab{a}}, \aap, 514,
  A67

\bibitem[{{Micha{\l}owski} et~al.(2008){Micha{\l}owski}, {Hjorth}, {Castro
  Cer{\'o}n}, \& {Watson}}]{michalowski08}
{Micha{\l}owski} M.J., {Hjorth} J., {Castro Cer{\'o}n} J.M., {Watson} D., 2008,
  \apj, 672, 817

\bibitem[{{Micha{\l}owski} et~al.(2009){Micha{\l}owski}, {Hjorth}, {Malesani}
  et~al.}]{michalowski09}
{Micha{\l}owski} M.J., {Hjorth} J., {Malesani} D., et~al., 2009, \apj, 693, 347

\bibitem[{{Micha{\l}owski} et~al.(2010{\natexlab{b}}){Micha{\l}owski},
  {Watson}, \& {Hjorth}}]{michalowski10smg4}
{Micha{\l}owski} M.J., {Watson} D., {Hjorth} J., 2010{\natexlab{b}}, \apj, 712,
  942

\bibitem[{{Morgan} \& {Edmunds}(2003)}]{morgan03}
{Morgan} H.L., {Edmunds} M.G., 2003, \mnras, 343, 427

\bibitem[{{Morgan} et~al.(2003){Morgan}, {Dunne}, {Eales}, {Ivison}, \&
  {Edmunds}}]{morgan03b}
{Morgan} H.L., {Dunne} L., {Eales} S.A., {Ivison} R.J., {Edmunds} M.G., 2003,
  \apjl, 597, L33

\bibitem[{{Nozawa} et~al.(2003){Nozawa}, {Kozasa}, {Umeda}, {Maeda}, \&
  {Nomoto}}]{nozawa03}
{Nozawa} T., {Kozasa} T., {Umeda} H., {Maeda} K., {Nomoto} K., 2003, \apj, 598,
  785

\bibitem[{{Perley} et~al.(2009){Perley}, {Bloom}, {Klein} et~al.}]{perley09}
{Perley} D.A., {Bloom} J.S., {Klein} C.R., et~al., 2009, \mnras, submitted,
  {\tt arXiv:0912.2999}

\bibitem[{{Petric} et~al.(2003){Petric}, {Carilli}, {Bertoldi}
  et~al.}]{petric03}
{Petric} A.O., {Carilli} C.L., {Bertoldi} F., et~al., 2003, \aj, 126, 15

\bibitem[{{Polletta} et~al.(2008){Polletta}, {Weedman}, {H{\"o}nig}
  et~al.}]{polletta08}
{Polletta} M., {Weedman} D., {H{\"o}nig} S., et~al., 2008, \apj, 675, 960

\bibitem[{{Priddey} \& {McMahon}(2001)}]{priddey01}
{Priddey} R.S., {McMahon} R.G., 2001, \mnras, 324, L17

\bibitem[{{Priddey} et~al.(2003){Priddey}, {Isaak}, {McMahon}, {Robson}, \&
  {Pearson}}]{priddey03}
{Priddey} R.S., {Isaak} K.G., {McMahon} R.G., {Robson} E.I., {Pearson} C.P.,
  2003, \mnras, 344, L74

\bibitem[{{Priddey} et~al.(2008){Priddey}, {Ivison}, \& {Isaak}}]{priddey08}
{Priddey} R.S., {Ivison} R.J., {Isaak} K.G., 2008, \mnras, 383, 289

\bibitem[{{Rho} et~al.(2008){Rho}, {Kozasa}, {Reach} et~al.}]{rho08}
{Rho} J., {Kozasa} T., {Reach} W.T., et~al., 2008, \apj, 673, 271

\bibitem[{{Rho} et~al.(2009){Rho}, {Reach}, {Tappe} et~al.}]{rho09}
{Rho} J., {Reach} W.T., {Tappe} A., et~al., 2009, \apj, 700, 579

\bibitem[{{Ridgway} et~al.(2001){Ridgway}, {Heckman}, {Calzetti}, \&
  {Lehnert}}]{ridgway01}
{Ridgway} S.E., {Heckman} T.M., {Calzetti} D., {Lehnert} M., 2001, \apj, 550,
  122

\bibitem[{{Riechers} et~al.(2007){Riechers}, {Walter}, {Carilli}, \&
  {Bertoldi}}]{riechers07}
{Riechers} D., {Walter} F., {Carilli} C., {Bertoldi} F., 2007, \apjl, 671, L13

\bibitem[{{Robson} et~al.(2004){Robson}, {Priddey}, {Isaak}, \&
  {McMahon}}]{robson04}
{Robson} I., {Priddey} R.S., {Isaak} K.G., {McMahon} R.G., 2004, \mnras, 351,
  L29

\bibitem[{{Sakon} et~al.(2009){Sakon}, {Onaka}, {Wada} et~al.}]{sakon09}
{Sakon} I., {Onaka} T., {Wada} T., et~al., 2009, \apj, 692, 546

\bibitem[{{Salpeter}(1955)}]{salpeter}
{Salpeter} E.E., 1955, \apj, 121, 161

\bibitem[{{Sandstrom} et~al.(2009){Sandstrom}, {Bolatto}, {Stanimirovi{\'c}},
  {van Loon}, \& {Smith}}]{sandstrom09}
{Sandstrom} K.M., {Bolatto} A.D., {Stanimirovi{\'c}} S., {van Loon} J.T.,
  {Smith} J.D.T., 2009, \apj, 696, 2138

\bibitem[{{Sibthorpe} et~al.(2009){Sibthorpe}, {Ade}, {Bock}
  et~al.}]{sibthorpe09}
{Sibthorpe} B., {Ade} P.A.R., {Bock} J.J., et~al., 2009, \apj, accepted {\tt
  arXiv:0910.1094}

\bibitem[{{Sloan} et~al.(2009){Sloan}, {Matsuura}, {Zijlstra} et~al.}]{sloan09}
{Sloan} G.C., {Matsuura} M., {Zijlstra} A.A., et~al., 2009, Science, 323, 353

\bibitem[{{Stratta} et~al.(2007){Stratta}, {Maiolino}, {Fiore}, \&
  {D'Elia}}]{stratta07b}
{Stratta} G., {Maiolino} R., {Fiore} F., {D'Elia} V., 2007, \apjl, 661, L9

\bibitem[{{Sugerman} et~al.(2006){Sugerman}, {Ercolano}, {Barlow}
  et~al.}]{sugerman06}
{Sugerman} B., {Ercolano} B., {Barlow} M., et~al., 2006, Science, 313, 196

\bibitem[{{Todini} \& {Ferrara}(2001)}]{todini01}
{Todini} P., {Ferrara} A., 2001, \mnras, 325, 726

\bibitem[{{Valiante} et~al.(2009){Valiante}, {Schneider}, {Bianchi}, \&
  {Andersen}}]{valiante09}
{Valiante} R., {Schneider} R., {Bianchi} S., {Andersen} A.C., 2009, \mnras,
  397, 1661

\bibitem[{{Veilleux} et~al.(2009){Veilleux}, {Kim}, {Rupke}
  et~al.}]{veilleux09}
{Veilleux} S., {Kim} D., {Rupke} D.S.N., et~al., 2009, \apj, 701, 587

\bibitem[{{Vlahakis} et~al.(2005){Vlahakis}, {Dunne}, \& {Eales}}]{vlahakis05}
{Vlahakis} C., {Dunne} L., {Eales} S., 2005, \mnras, 364, 1253

\bibitem[{{Walter} et~al.(2003){Walter}, {Bertoldi}, {Carilli}
  et~al.}]{walter03}
{Walter} F., {Bertoldi} F., {Carilli} C., et~al., 2003, \nat, 424, 406

\bibitem[{{Walter} et~al.(2004){Walter}, {Carilli}, {Bertoldi}
  et~al.}]{walter04}
{Walter} F., {Carilli} C., {Bertoldi} F., et~al., 2004, \apjl, 615, L17

\bibitem[{{Walter} et~al.(2009){Walter}, {Riechers}, {Cox} et~al.}]{walter09}
{Walter} F., {Riechers} D., {Cox} P., et~al., 2009, \nat, 457, 699

\bibitem[{{Wang} et~al.(2007){Wang}, {Carilli}, {Beelen} et~al.}]{wang07b}
{Wang} R., {Carilli} C.L., {Beelen} A., et~al., 2007, \aj, 134, 617

\bibitem[{{Wang} et~al.(2008{\natexlab{a}}){Wang}, {Carilli}, {Wagg}
  et~al.}]{wang08}
{Wang} R., {Carilli} C.L., {Wagg} J., et~al., 2008{\natexlab{a}}, \apj, 687,
  848

\bibitem[{{Wang} et~al.(2008{\natexlab{b}}){Wang}, {Wagg}, {Carilli}
  et~al.}]{wang08b}
{Wang} R., {Wagg} J., {Carilli} C.L., et~al., 2008{\natexlab{b}}, \aj, 135,
  1201

\bibitem[{{Wang} et~al.(2010){Wang}, {Carilli}, {Neri} et~al.}]{wang10}
{Wang} R., {Carilli} C.L., {Neri} R., et~al., 2010, \apj, 714, 699

\bibitem[{{Wesson} et~al.(2009){Wesson}, {Barlow}, {Ercolano}
  et~al.}]{wesson09}
{Wesson} R., {Barlow} M.J., {Ercolano} B., et~al., 2009, \mnras, 403, 474

\bibitem[{{Wilson} \& {Batrla}(2005)}]{wilson05}
{Wilson} T.L., {Batrla} W., 2005, \aap, 430, 561

\bibitem[{{Wu} et~al.(2009){Wu}, {Vanden Bout}, {Evans}, \& {Dunham}}]{wu09}
{Wu} J., {Vanden Bout} P., {Evans} N., {Dunham} M., 2009, \apj, 707, 988

\bibitem[{{Zafar} et~al.(2010){Zafar}, {Watson}, {Malesani} et~al.}]{zafar10}
{Zafar} T., {Watson} D.J., {Malesani} D., et~al., 2010, \aap, 515, A94

\bibitem[{{Zhukovska} et~al.(2008){Zhukovska}, {Gail}, \&
  {Trieloff}}]{zhukovska08}
{Zhukovska} S., {Gail} H.P., {Trieloff} M., 2008, \aap, 479, 453

\end{thebibliography}
